# Using coherent X-rays to directly measure the propagation velocity of defects during thin film deposition

Jeffrey G. Ulbrandt [1], Meliha G. Rainville [2], Christa Hoskin [2], Suresh Narayanan [3], Alec R. Sandy [3], Hua Zhou [3], Karl F. Ludwig, Jr. [2,4] and Randall L. Headrick [1*]

[1] *Department of Physics and Materials Science Program, University of Vermont, Burlington, Vermont 05405 USA*

[2] *Division of Materials Science and Engineering, Boston University, Boston, Massachusetts 02215 USA*

[3] *Advanced Photon Source, Argonne National Lab, Argonne, IL, 60439 USA*

[4] *Department of Physics, Boston University, Boston, Massachusetts 02215 USA*

[*]Email: rheadrick@uvm.edu

## Abstract

The properties of artificially grown thin films are often strongly affected by the dynamic relationship between surface growth processes and subsurface structure. Coherent mixing of X-ray signals promises to provide an approach to better understand such processes. Here, we demonstrate the continuously variable mixing of surface and bulk scattering signals during real-time studies of sputter deposition of a-Si and a-WSi$_2$ films by controlling the X-ray penetration and escape depths in coherent grazing incidence small angle X-ray scattering (Co-GISAXS). Under conditions where the X-ray signal comes from both the growth surface and the thin film bulk, oscillations in temporal correlations arise from coherent interference between scattering from stationary bulk features and from the advancing surface.  We also observe evidence that elongated bulk features propagate upward at the same velocity as the surface.  Additionally, a highly surface sensitive mode is demonstrated that can access the surface dynamics independently of the subsurface structure.

A key objective for understanding surface dynamics during thin film growth is the ability to monitor nanometer-scale surface fluctuation dynamics in real time[1-3]. These fluctuations of roughness and density occur on timescales that rarely exceed a few seconds, and take place in environments that are inaccessible to most high spatial resolution probes. For example, scanning probe microscopy is widely used to study interfacial reactivity in non-vacuum environments[4], but is limited by inability to probe surfaces in real time during deposition; electron microscopy is mainly limited to high vacuum environments and low magnetic fields[5,6]. X-rays have the potential to overcome these challenges due to their highly penetrating nature and sensitivity to nanometer-scale features. Observation of subsurface structures in real time during film growth appears to be even more challenging, and has rarely been attempted[7]. Bulk signals are sometimes observed as unwanted background in grazing incidence surface X-ray scattering experiments, but there have been few attempts to quantitatively understand the features responsible for such signals[8,9].

Interaction of surfaces with nanometer-scale buried defects and formation of bulk defects at a growing surface are integral to many industrial processes. For example, misfit dislocations nucleate at free surfaces and buried interfaces in strained layer epitaxial growth of layers for photonic devices[10], motion and ordering of oxygen vacancies in complex oxide materials for ferroelectric memory depend on the surface conditions during growth[11-13], and voids in electrochemically deposited layers used for interconnects in electronic circuits are introduced by surface processes during deposition[14].

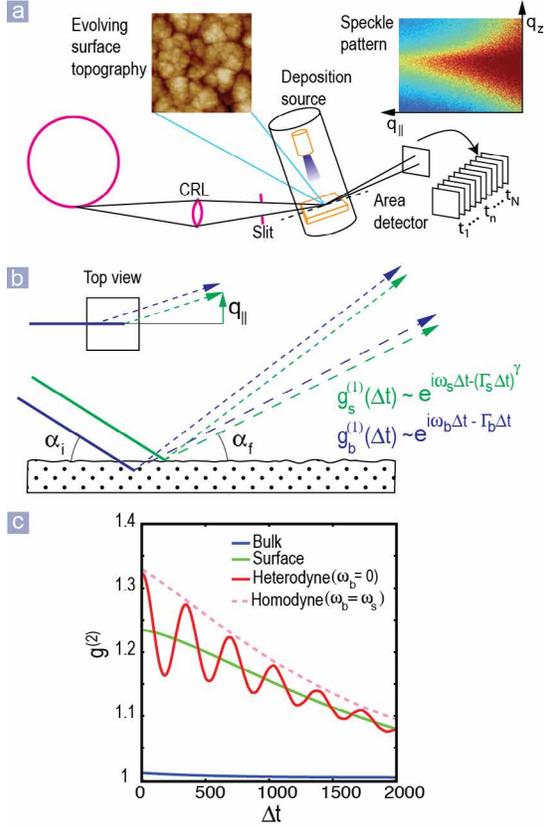

**Figure 1 Schematic of the experiment and coherent mixing effects.** (**a**) X-rays from the synchrotron source are focused by a compound refractive lens (CRL) and a collimating slit system into an ultra-high vacuum sample enclosure. An amorphous thin film is deposited, which causes the surface to advance at the growth velocity, and also induces random fluctuations in the surface roughness. Scattered coherent X-rays form a speckle pattern that corresponds to the detailed configuration of the surface, which is recorded versus time by a high-resolution photon sensitive X-ray area detector. (**b**) In addition to scattering from the surface (green lines and equation), the X-rays penetrate beneath the surface and may be scattered by density variations in the bulk of the film (blue). The functions $g_s^{(1)}(\Delta t)$ and $g_b^{(1)}(\Delta t)$ correspond to the intermediate scattering functions for surface and bulk contributions respectively. (**c**) The two signals interfere coherently creating temporal correlations in the speckle pattern that can oscillate if the frequencies $\omega_s$ and $\omega_b$ of the two components differ. Note that the surface component $g_s^{(1)}(\Delta t)$ advances in phase with a frequency $\omega_s$ that is determined by the film growth velocity, and the bulk component $g_b^{(1)}(\Delta t)$ may be in-phase with the surface (homodyne mixing mode) or advance with a different phase (heterodyne mode), depending on the nature of the features responsible for the bulk scattering. The second order correlation function $g^{(2)}(\Delta t)$ can be extracted directly from intensity data, as described in the main text.

The use of X-ray scattering techniques to probe *in situ* real-time processes has largely been restricted to well-ordered crystalline structures and to statistical averages of disorder due to limitations in the spectral brightness of X-ray sources. A fundamental limitation in this regard is the coherence length of X-rays, usually <1 µm, which imposes an averaging over many coherence volumes contained within X-ray beams of typical dimensions[15-17]. However, recent advances in high spectral brightness sources, along with parallel advances in X-ray area detectors has led to a new frontier in X-ray scattering through techniques such as X-ray photon correlation spectroscopy (XPCS)[18-24] that can follow the natural thermal fluctuations in condensed matter systems[15,25]. Here, we describe a new application of coherent X-rays that extends scattering studies to observation of local fluctuation dynamics during film growth via XPCS, and also provides a sensitive measure of the relative propagation velocity of surface and subsurface features. This technique opens up possibilities for studies of surfaces[16], interfaces[26,27], and bulk defects[8,9], such as for crystal growth in the step-flow mode, where monolayer steps propagate with a well-defined velocity and direction[28,29].

When a coherent beam of light falls on an object with any type of disorder, static or dynamic, the scattered light forms a speckle pattern composed of an apparently random array of bright spots (Fig 1a). If the different parts of the object fluctuate in time, the speckle pattern also fluctuates on the same timescale due to the changing pattern of interference between the scattered waves. In this case where dynamics are present, XPCS can characterize fluctuation time scales as a function of length scale via the X-ray wave vector transfer $q$. XPCS has previously been utilized for studies of thermal fluctuations on surfaces such as capillary waves on liquid surfaces and polymer film surfaces.[1,3,30,31] Employing heterodyne measurements by mixing in a reference signal can increase the signal-to-noise ratio and also opens the possibility of obtaining phase information. Heterodyne mixing from bulk samples has been observed up to a wave vector transfer of $q \sim 0.02$ Å$^{-1}$ using an external random scatterer as a static reference.[32] Heterodyne mixing of surface scattering has been observed at relatively small values of the in-plane wave vector transfer, $q_{||}$, on the order of $10^{-6} - 10^{-5}$ Å$^{-1}$ (i.e. rather large length scales) due to a fortuitous overlap between the tails of the specular reflected beam and the non-specular surface scattering. However, coherent mixing of surface scattering with a reference wave at larger scattering vector (smaller length scales) has not been demonstrated. Here, we report the observation of a new phenomenon -- coherent mixing of the bulk and surface waves (Fig. 1b,c) -- that provides access to the relative phases of the scattered signals up to $q_{||} \sim 0.12$ Å$^{-1}$. The strategy that we employ is to use the surface advancing at a constant growth velocity to produce a quasi-static reference wave in order to deduce whether bulk defects propagate along with the surface as they form (homodyne mixing case) or whether they form localized features that do not propagate (heterodyne mixing case).

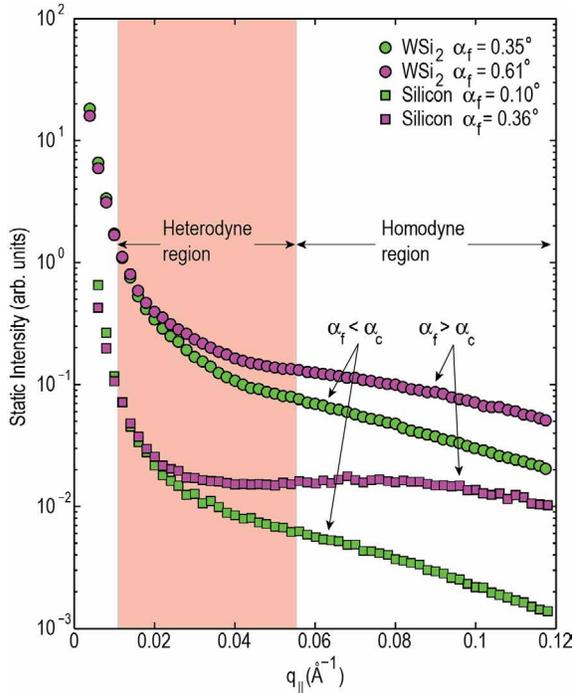

**Figure 2 Static intensities during steady state growth.** Data averaged over speckles for exit angles $\alpha_f$ above and below $\alpha_c$, the critical angle for total external reflection as a function of the in-plane component of the scattering vector **q**. Two sets of curves are shown: for a-Silicon film (squares) and a-WSi$_2$ film (circles). The intensity at angles below the critical angle is lower because the X-ray escape depth is greatly reduced, and the difference between the intensities at the two different exit angles gives a qualitative indication of the bulk contribution. Note that the critical angles measured for WSi$_2$ and Si with 7.35 keV X-rays are approximately $\alpha_c$ = 0.40° and 0.21° respectively. The incidence angles are $\alpha_i$ = 0.39° and $\alpha_i$ = 0.26° respectively.

## Results

We apply this novel approach to studies of amorphous thin films during magnetron sputter deposition by monitoring the grazing incidence small angle X-ray scattering (GISAXS) speckle pattern in real time (Fig. 1). The angle of incidence is chosen to be close to the critical angle for total external reflection, or slightly above it so that the signal escape depth can be varied over a wide range by changing the exit angle between the surface and the detected X-rays. Before discussing the results in detail, we describe three principles of the measurement technique: (i) preparing a steady-state growth surface, (ii) extracting correlation decay lineshapes from a sequence of images, and (iii) controlling the degree of mixing between surface and bulk waves by varying the exit angle.

An important aspect of this measurement is that the surface is prepared in a state of stationary surface dynamics where the average properties of the surface such as roughness are unchanging but local fluctuations occur as long as the deposition continues. We find that amorphous surfaces often obey the Family-Vicsek scaling relation for surface growth[33,34]:

$$w(L,t) \sim L^\alpha f\left(\frac{t}{L^z}\right) \quad (1)$$

where $w(L,t)$ is the interface width due to roughness, $L$ is the system size or lateral length scale, $z$ is the dynamic growth exponent and $\alpha$ is the roughness exponent. The scaling function $f(u)$ satisfies $f(u) \to 1$ for $u \to \infty$, and so the surface width approaches a steady state value within the range of length scales accessible to the experiment. This is verified by monitoring the static GISAXS intensity averaged over the speckles (Fig. 2 and Supplementary Figs. S3 and S4). Information about local fluctuations is contained in the time-time correlation function[35],

$$<h(q_{||},t_1)h(q_{||},t_2)> \sim g_{ss}(q_{||}{}^z|t_1 - t_2|) \quad (2)$$

where $h(q_{||},t)$ is the Fourier component of the surface amplitude at wave vector $q_{||}$ and time $t$. This expression is valid for $t_1, t_2 \to \infty$ and $|t_1 - t_2|$ finite, i.e. the so-called steady-state regime. Also, $\lim_{x \to \infty} g_{ss}(x) = 0$, so Equation 2 implies that the correlations decay with a time constant $\tau_s(q_\|) \sim q_\|^{-z}$. We find that for conditions used in this work, the surface time constants are relatively long (compared to, e.g. the expected timescales for surfaces during epitaxial growth), ~100 seconds or larger (Fig. 3 and Supplementary Tables 1&2).

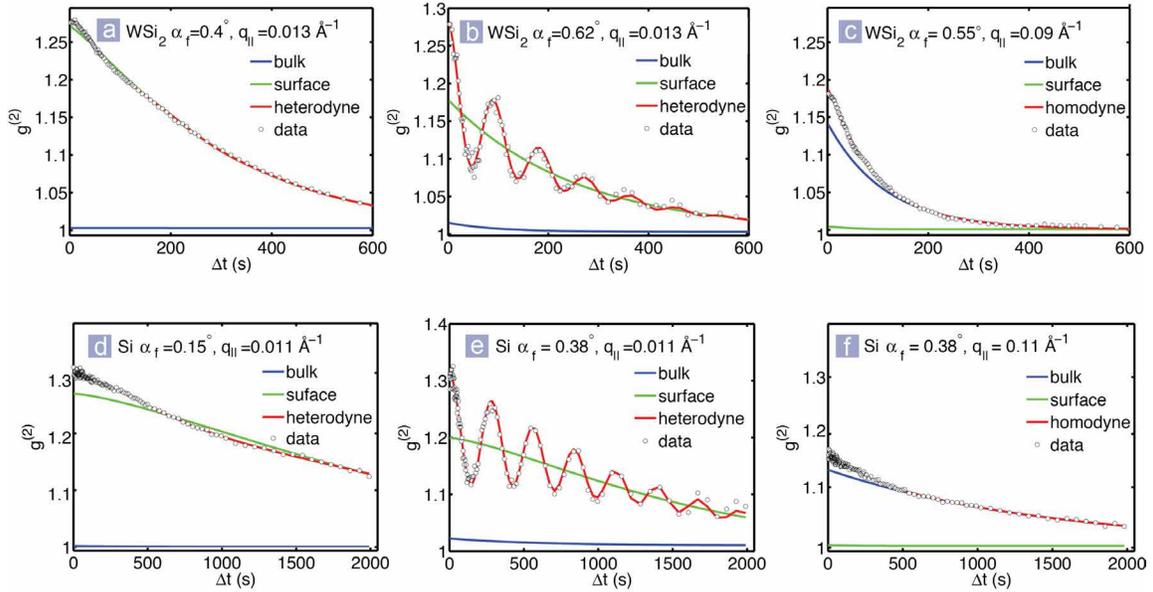

**Figure 3 Identification of heterodyne and homodyne mixing.** Examples of temporal correlation data and fitting results for (**a**-**c**) WSi$_2$ and (**d**-**f**) Si are shown. The surface component dominates at grazing exit angles below the critical angle and the heterodyne curve calculated from Equation 9 is almost indistinguishable from the surface homodyne curve calculated from Equation 6 (**a** and **d**). In contrast, the bulk component becomes significant enough to produce a strong heterodyne effect for exit angles above the critical angle (**b** and **e**). At larger $q_\|$, the bulk component becomes dominant, but no heterodyne oscillations are observed for any exit angle (**c** and **f** show results for high $q_\|$ and large exit angle). This is interpreted as the bulk and surface contributions having the same frequency, i.e. $\omega_b = \omega_s$, in the 2-wave homodyne mixing mode.

A second important aspect of this method is the evaluation of the surface correlation decay time constant $\tau_s(q_\|)$ from experimental data and comparison to theoretical models. We utilize standard data reduction methods of XPCS by relating the measured intensity versus time $I(q,t)$ to the intensity autocorrelation function

$$g^{(2)}(\boldsymbol{q},t) = \frac{\langle I(\boldsymbol{q},t')I(\boldsymbol{q},t'+t)\rangle}{\langle I(\boldsymbol{q})\rangle^2} \qquad (3)$$

The above equation can be decomposed into a simpler product of correlation functions of electric fields rather than intensities

$$g^{(2)}(\boldsymbol{q},t) = 1 + \beta(\boldsymbol{q})|F(\boldsymbol{q},t)|^2 \qquad (4)$$

where $F(\boldsymbol{q},t) = g^{(1)}(\boldsymbol{q},t)/g^{(1)}(\boldsymbol{q},0)$ is the normalized intermediate scattering function with $g^{(1)}(\boldsymbol{q},t) \sim \langle E(\boldsymbol{q},t')E^*(\boldsymbol{q},t'+t)\rangle$, and $\beta(\boldsymbol{q})$ is the optical contrast factor. The intermediate scattering function is related to density-density variations in the sample, and in the case of GISAXS the surface scattering is related to variations in the height of the surface through $g_s^{(1)}(\boldsymbol{q},t) \sim \langle h(q_{||},t')h^*(q_{||},t'+t)\rangle$. It follows from our discussion of the statistical properties of growing surfaces that the intermediate scattering function takes the form

$$g_s^{(1)}(\mathbf{q},t) \sim \exp\{i\omega_s t - [\Gamma_s(q_{||})t]^\gamma\} \qquad (5)$$

where $\omega_s = q_z v$ is the product of the component of the momentum transfer perpendicular to the growing surface and the growth velocity $v$, and $\Gamma_s(q_{||}) = 1/\tau_s(q_{||})$. This form matches closely to the form describing capillary waves on liquid surfaces[30], except that capillary waves propagate in the plane of the surface and so the phase depends on the in-plane component of the wave vector transfer, while in the case of surface growth the velocity is normal to the surface. The stretching exponent $\gamma$ takes into account the possibility that the equation of motion of the surface is non-linear, i.e. that it may include terms such as $(\nabla h)^2$, in which case the decay of the correlations does not have to be a simple exponential. In the absence of any optical mixing, i.e. neglecting the bulk contribution to the scattering, the theoretical expression for the intensity autocorrelation function becomes

$$g_s^{(2)}(\boldsymbol{q}, t) = 1 + \beta(\boldsymbol{q}) \exp\{-2[\Gamma_s(q_{\parallel})t]^{\gamma}\} \qquad (6)$$

Note that the phase information is lost in the single-wave homodyne detection scheme. This mode is achievable in GISAXS by varying the incidence and detection angles $\alpha_i$ and $\alpha_f$ of the X-rays with respect to the surface. Fig. 3a illustrates a case where the incidence and exit angles are both less than the critical angle for total external reflection $\alpha_c$, and the decay of the correlations is consistent with Equation 6. Curve fitting results show that surface correlations decay with $\gamma \approx 1.2 - 1.7$, indicating a compressed exponential line shape (Supplementary Tables S1 and S2).

The third, and most novel aspect of this method is heterodyne mixing. We vary the incidence or exit angle to control the X-ray penetration and escape depths in order to control the amount of mixing between the surface and bulk waves. Note that the bulk signal is *not* assumed to be entirely static, since features formed at or just beneath a growing surface will no longer contribute to the signal after they become deeply buried. In contrast, previous heterodyne mode XPCS methods have relied on a perfectly static reference wave[1,30,32,36,37]. Our approach here is that it is not necessary to have a perfectly static reference wave, rather we have a symmetric situation where either wave can be considered to be the reference. This does somewhat complicate the analysis and fitting since in general there are more unknown parameters to determine. However, tuning the mixing ratio by varying the X-ray penetration and escape depths over a wide range provides a powerful method for decoupling the two signals or mixing them in almost any ratio desired. Following the discussion of surface correlations above, our trial form for the intermediate scattering function for the bulk wave is

$$g_b^{(1)}(\mathbf{q}, t) \sim \exp\{i\omega_b t - \Gamma_b t\} \qquad (7)$$

If we assume that randomly distributed bulk features do not segregate to the surface, then for a given angle of incidence the maximum time constant at large exit angles should be $\tau_{b,max} \approx \Lambda/v$, where $\Lambda$ is the penetration depth of the incident X-rays and $v$ is the growth velocity. For example in the case of a Silicon surface with 7.35 keV X-rays incident at 0.26°, the penetration depth is about 1500 Å, so that with a growth velocity of 0.57 Å/s, we estimate $\tau_{b,max} \approx 2500$ s (Fig. 5b). Moreover, the assumption that the bulk features do not segregate also implies that they have zero velocity, so that $\omega_b = 0$.

The intensity autocorrelation function for coherent mixing of two waves with intensities $I_s$ and $I_b$ evaluates to

$$\langle I(\mathbf{q}, t')I(\mathbf{q}, t'+t)\rangle - \langle I(\mathbf{q})\rangle^2$$
$$= I_s^2\left[g_s^{(2)}(\mathbf{q}, t) - 1\right] + I_b^2\left[g_b^{(2)}(\mathbf{q}, t) - 1\right] + 2\beta I_b I_s \mathcal{R}e\left[g_s^{(1)}(\mathbf{q}, t)g_b^{(1)*}(\mathbf{q}, t)\right] \qquad (8)$$

Note that we can recover the form for a static reference wave by setting one of the $g^{(1)}(\mathbf{q}, t)$ functions in Equation 8 to unity and the corresponding $g^{(2)}(\mathbf{q}, t)$ equal to $1 + \beta$. Upon inserting the intermediate scattering functions for both waves from Equations 5 and 7, we have

$$\langle I(\boldsymbol{q},t')I(\boldsymbol{q},t'+t)\rangle \sim 1 + I_s^2 \exp\{-2[\Gamma_s(q_\parallel)t]^\gamma\} + I_b^2\exp(-2\Gamma_b t)$$

$$+ 2I_b I_s \cos\{(\omega_s - \omega_b)t\} \exp\{-[\Gamma_s(q_\parallel)t]^\gamma - \Gamma_b t\} \quad (9)$$

Equation 9 is our master equation that can describe the correlations of any combination of surface and bulk waves with different intensities, time constants, and growth velocities. Note that the phase information is included in the last term, and that in the special case $\omega_b = 0$, heterodyning is observed and the correlation function oscillates with a period

$$T_{HD} = \frac{2\pi}{q_z v} \quad (10)$$

while for the case $\omega_b = \omega_s$ there will be no oscillations. We refer to the latter case as *homodyne mixing* to distinguish it from single-wave homodyne. Fig. 1c illustrates both heterodyne and homodyne mixing.

The measurements discussed above were carried out over a range of $q_\parallel$ up to 0.12 Å$^{-1}$ and $q_z$ up to ~0.03 Å$^{-1}$ for two different amorphous thin film systems, WSi$_2$ and Silicon. We confirmed that the averaged intensity was unchanging ~2000 s after the start of each deposition (Supplementary Figs. S3 and S4). Post-growth atomic force microscopy measurements showed surface roughness of 4-6 nm under the conditions used here. Fig. 2 shows evidence for bulk scattering since the speckle-averaged intensity for exit angles above the critical angle is larger than for below the critical angle. The bulk signal also converges to a steady state during the deposition due to the limited penetration depth of the X-rays at the grazing angles used in the experiment, as discussed above.

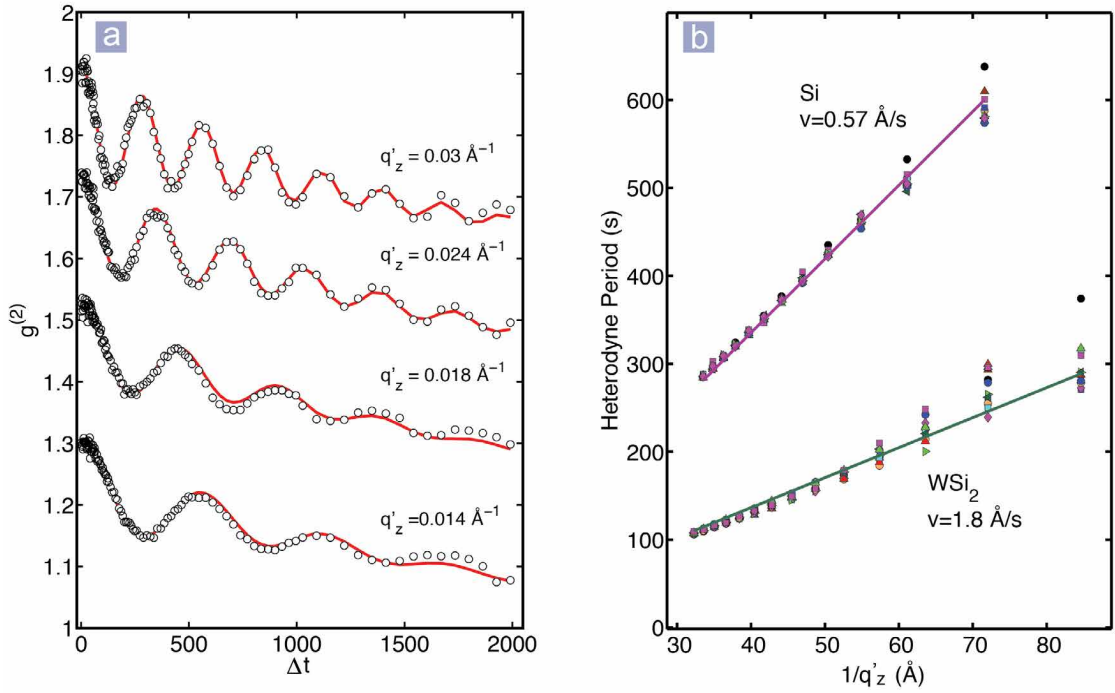

**Figure 4 Measurement of the surface velocity from the heterodyne period.** (**a**) Examples of heterodyne oscillations for Si at several $q_z'$ with $q_\parallel$ = 0.011 Å$^{-1}$. The red lines are fitted curves using the heterodyne model. Each curve is displaced from the one underneath it by 0.2. (**b**) The heterodyne period is found to be linear with $1/q_z'$, and independent of the in-plane component of $\boldsymbol{q}$. Data points are for several values of $q_\parallel$ between 0.006 - 0.026 Å$^{-1}$ and 0.022 - 0.042 Å$^{-1}$ for Si and WSi$_2$ respectively. Lines are the calculated heterodyne periods for the surface growth velocities indicated in the figure.

Correlation curves were calculated from the x-ray data using Equation 3, and were fitted using Equation 9 for each dynamic $q_\parallel$-$q_z$ mask region. The resulting correlation curves (Fig. 3) exhibit strong heterodyne mixing at low $q_\parallel$, up to about 0.05 Å$^{-1}$ in both cases, but only for exit angles above the critical angle (Fig. 3 b,e). Fig. 4 shows the heterodyne period extracted from the fitting results. In order to calculate $q_z' = k(\sin\alpha_f' + \sin\alpha_i')$, we use corrected incidence and exit angles $\alpha_i' = (\alpha_i^2 - \alpha_c^2)^{1/2}$ and $\alpha_f' = (\alpha_f^2 - \alpha_c^2)^{1/2}$. Fig. 4 shows that the heterodyne period varies linearly with $1/q_z'$, and does not vary systematically with $q_{||}$. Moreover, both growth velocities are in agreement with quartz crystal microbalance (QCM) calibration and post-deposition cross sectional Scanning Electron Microscopy measurements of film thickness within 20%. In order to make the QCM results agree with the measured growth velocities perfectly, we need to assume that the actual film density is less than the nominal bulk density (2.33 and 9.30 g/cm³ for Si and $WSi_2$, respectively). This is not unreasonable under the conditions used. For example, there have been reports of a density deficit, which increases with sputtering pressure in amorphous metal and semiconductor films[38]. It is notable that many transmission electron microscopy studies have been carried out that show ~100 Å voids in a broad range of thin films deposited by various methods, such as sputter deposition, thermal evaporation, and electroplating[14]. In addition, there have been reports of diffuse scattering in small angle electron diffraction patterns of amorphous Si thin films that were attributed to voids[39]. Thus voids or other defects are natural bulk scatterers to provide heterodyne interference.

The picture described above is appealing, but incomplete since we observe that the heterodyne oscillations disappear for $q_{||}$ > 0.05 Å⁻¹ although it is clear from Fig. 2 that the bulk signal is still present at the highest $q_{||}$. Fig. 5 shows that there are still two components present in the correlation decay plot at high $q_{||}$, and they can be distinguished in two ways: (i) the surface and bulk signals have different time constants, and (ii) surface and bulk wave intensities vary in a systematic way as a function of exit angle. The results show unambiguously that there are still two waves undergoing optical mixing. The lack of heterodyne oscillations suggests that there is a population of defects with very different characteristics (e.g. propagation or segregation velocity) compared to the features that produce the heterodyne signal at lower $q_{\parallel}$.

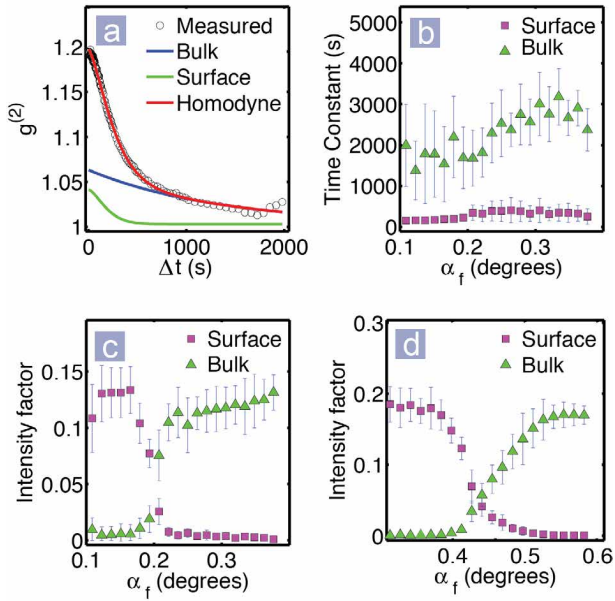

**Figure 5 Control of the homodyne mixing ratio by varying the exit angle.** (a-c) Summary of Silicon fitting at $q_{\parallel}$ = 0.1 Å$^{-1}$. (a) Correlation versus delay time at $\alpha_f$ = 0.21°. The optical mixing is modeled as homodyne in this region of $q_{\parallel}$ due the complete absence of heterodyne oscillations. Note that at this particular combination of $q_{\parallel}$ and exit angle the surface and bulk wave intensity factors ($I_s$ and $I_b$ in Equation 9) are nearly equal. (b) Time constants of surface and bulk contributions averaged over $q_{\parallel}$ = 0.10 to 0.12 Å$^{-1}$. The bulk and surface components are distinguishable since they have significantly different time constants. (c) Variation of surface and bulk wave intensities with exit angle. The surface component dominates below the critical angle, while the bulk component dominates above it. (d) Surface and bulk wave intensities for WSi$_2$ averaged over $q_{\parallel}$ = 0.08 to 0.10 Å$^{-1}$. The intensities cross at a higher angle compared to (c) because the critical angle is larger for WSi$_2$ than for Si.

We model this phenomenon as a bulk wave with a phase that advances with the surface so that $\omega_b = \omega_s$. In this case, the heterodyne term in Equation 9 is still present, but it does not vary with time and so there are no oscillations; this is the homodyne mixing mode. It is known that a wide variety of thin films (both amorphous and crystalline) form in a columnar morphology[40]. These features are readily observed with scanning electron microscopy, and for our films the column widths are several thousand angstroms (Supplementary Fig. S5). However, this is much too large to explain the bulk signal that we observe. Scanning tunneling microscopy experiments on Si sputter deposited onto highly oriented pyrolytic graphite have found bundles of much smaller nanowires with diameters in the range of 30 – 70 Å and at least 1000 Å long[41] This type of nanocolumnar structure with the long axis of the nanocolumns oriented perpendicular to the film surface has been suggested to be generally present in thin films prepared under low mobility conditions[42]. The columns are defined by a network of elongated voids or pores at their boundaries.[43-45] These features naturally produce scattering with a phase that advances with the surface under conditions where the vertical correlation length of the defects is comparable to or larger than the escape depth of the X-rays. Supplementary Tables 1&2 show that $\tau_b$ increases at large $q_{||}$. Multiplying these values by the growth velocities yields vertical correlation lengths of ~400 Å and ~1700 Å for $WSi_2$ and Si, respectively. The behavior of this second population of defects is thus consistent with elongated features that grow and interact with the surface, with lengths similar to those described previously.[41,46]

## Conclusions

Coherent mixing of surface and bulk X-ray scattering waves provides a powerful way to recover relative phase information during thin film deposition. We have applied the effect to measure the dynamics of nanoscale surface and sub-surface features that are not readily distinguishable from each other in conventional X-ray scattering. The results reveal surprisingly rich insights into thin film growth dynamics and defect formation. We conclude that there are two defect populations: compact void-like features forming near the surface that are buried during deposition, and a second population of elongated column-like features. The void scattering mixes with the surface scattering to produce a heterodyne signal, with oscillations arising from the relative motion of the growing surface with respect to the defects. The oscillation frequency corresponds well with the surface growth velocity, implying that the voids do not segregate ( $v_b$ = 0). In contrast, the scattering from the sides of the nanocolumns mixes with the surface scattering to produce a two-wave homodyne signal, since these features propagate upward in coincidence with the surface. A highly surface sensitive mode is also demonstrated, where the surface dynamics itself is accessed independently of the subsurface structure. The ability to monitor these fundamental processes using Coherent GISAXS represents an important step forward in elucidating the nanoscale mechanisms underlying thin film deposition processes.

# Methods

The experiments were carried out at beamline 8-ID-I at the Advanced Photon Source (APS) at Argonne National Laboratory. An X-ray wavelength of 1.69 Å (E = 7.35 keV) was selected using a double bounce Ge(111) monochromator with a bandwidth of $\Delta\lambda/\lambda \sim 3\times10^{-4}$. The X-ray beam was focused to 20 μm (H) by 4 μm (V) at the sample through a compound refractive lens and collimating slits, with a flux of $\sim 7\times10^{10}$ ph/s. A direct detection charge-coupled-device with 20 μm pixels (Princeton Instruments LCX-1300) was placed 4 m from the sample.

A custom stainless steel vacuum chamber with Beryllium windows was constructed for this experiment. The chamber is pumped via a turbo-molecular pump with magnetically levitated bearings (Edwards STP-301), with a 65 lb. vibration isolator installed on the backing line, and backed by a scroll pump. The sample is held on a sample stage in vertical reflection geometry with no in-vacuum motions. The entire chamber is rotated about an axis that passes through the sample surface via a 2-circle segment (Huber 5203), which is part of the standard beamline setup. See Supplementary Fig. S6 for a photograph of the system installed at APS 8-ID-I.

A downward facing water-cooled sputter gun (Meivac) capable of holding 2" diameter targets is used as the deposition source. It is placed at a distance of 100 mm above the substrate surface in normal incidence. $WSi_2$ and Si sputtering targets purchased from Kurt Lesker Corp. are bonded to copper backing plates. Substrates are either pieces of Silicon wafers (for the Si depositions) or Silicon wafers with a 500 nm thermal oxide (for the WSi$_2$ depositions). A sputtering power of 20 W (for the Si depositions) or 25W (for the WSi$_2$ depositions) is produced by a DC power supply (Advanced Energy MDX500). The Argon pressure during sputtering is between 10 and 16 mTorr. This pressure range is chosen because it is above the roughening transition pressure for WSi$_2$ (~6 mTorr)[47]. The sample stage was replaced with a quartz crystal microbalance for calibration of deposition rates. Calibration runs were done in the same chamber both before and after the X-ray experiments, and were found to be reproducible within 3%, which indicates that changes in the deposition rates during the experiments due to erosion of the targets or other factors was minimal. Several post-deposition measurements were performed on the films, including Atomic Force Microscopy to characterize the surface roughness and cross-section Scanning Electron Microscopy to image the microstructure in the bulk of the films.

Data collection scans are performed at a fixed angle of incidence. The detector intercepts a total range of q$_{||}$ ~ 0.025 Å$^{-1}$ during each scan, so 6 overlapping regions were used to reach the full range of $q_{||}$, 0.003 to 0.12 Å$^{-1}$ We define $q_{||}$ as being the component of the wave vector transfer $\boldsymbol{q} = \boldsymbol{k}_f - \boldsymbol{k}_i$ in the plane of the surface, while the perpendicular component is denoted as $q_z$. All scans are performed on each sample during a single long continuous deposition in order to ensure that the sample surface is maintained in steady-state conditions. Each scan consists of 50 dark images, followed by 1024 images with 2 seconds integration. Note that the first scan for each sample during the transient period of surface development is not used for the steady-state analysis. See Supplementary Figs. S3 and S4 for examples of the transient behavior.

XPCSGUI is a custom MATLAB based analysis package for XPCS data sets. XPCSGUI was used to define $q_{||}$-$q_z$ mask regions and to compute intensity autocorrelations and two-time correlations. Fitting of $g^{(2)}(t)$ intensity autocorrelation curves is accomplished with a least-squares Levenberg-Marquardt minimization with a fitting function based on Equation 9. See Supplementary Tables S1 and S2 for detailed fitting results.

## ACKNOWLEDGEMENTS

We thank Ray Ziegler for beamline support. RH and JU were supported by the U.S. Department of Energy (DOE) Office of Science, Office of Basic Energy Sciences (BES) under DE-FG02-07ER46380; CH, KL, and MR were supported by DOE BES grant DE-FG02-03ER46037. This research used resources of the Advanced Photon Source (APS), a U.S. DOE Office of Science User Facility operated for the DOE Office of Science by Argonne National Laboratory under Contract No. DE-AC02-06CH11357.

## AUTHOR CONTRIBUTIONS

All authors participated in the experimental work at the APS. RH, CH, and MR were chiefly responsible for the subsequent processing and analysis of the data with significant contributions from SN. RH and JU designed and constructed the growth chamber. RH and KL devised the new experimental methods and directed the project. All authors participated in preparing the manuscript.

## REFERENCES

1   Sinha, S. K., Jiang, Z. & Lurio, L. B. X-ray Photon Correlation Spectroscopy Studies of Surfaces and Thin Films. *Advanced Materials* **26**, 7764-7785, doi:10.1002/Adma.201401094 (2014).

2   Pierce, M. S., Chang, K. C., Hennessy, D., Komanicky, V., Sprung, M., Sandy, A. & You, H. Surface X-Ray Speckles: Coherent Surface Diffraction from Au(001). *Physical Review Letters* **103**, 165501, doi:10.1103/Physrevlett.103.165501 (2009).

3   Kim, H., Ruhm, A., Lurio, L. B., Basu, J. K., Lal, J., Lumma, D., Mochrie, S. G. J. & Sinha, S. K. Surface dynamics of polymer films. *Physical Review Letters* **90**, 068302, doi:10.1103/Physrevlett.90.068302 (2003).

4   Ohnesorge, F. & Binnig, G. True Atomic-Resolution by Atomic Force Microscopy through Repulsive and Attractive Forces. *Science* **260**, 1451-1456, doi:10.1126/Science.260.5113.1451 (1993).

5   Bauer, E. Low-Energy-Electron Microscopy. *Reports on Progress in Physics* **57**, 895-938, doi:10.1088/0034-4885/57/9/002 (1994).

6   Leslie, C., Landree, E., Collazo-Davila, C., Bengu, E., Grozea, D. & Marks, L. D. Electron crystallography in surface structure analysis. *Microscopy Research and Technique* **46**, 160-177, doi:10.1002/(Sici)1097-0029(19990801)46:3<160::Aid-Jemt2>3.0.Co;2-# (1999).

7   Bein, B., Hsing, H.-C., Callori, S. J., Sinsheimer, J., Chinta, P. V., Headrick, R. L. & Dawber, M. Rapid in-situ x-ray diffraction during the growth of ferroelectric superlattices. (arXiv:1502.07632, 2015).

8   Pfeiffer, F., Zhang, W. & Robinson, I. K. Coherent grazing exit x-ray scattering geometry for probing the structure of thin films. *Applied Physics Letters* **84**, 1847-1849, doi:10.1063/1.1669061 (2004).

9   Ferguson, J. D., Kim, Y., Kourkoutis, L. F., Vodnick, A., Woll, A. R., Muller, D. A. & Brock, J. D. Epitaxial Oxygen Getter for a Brownmillerite Phase Transformation in Manganite Films. *Advanced Materials* **23**, 1226, doi:10.1002/Adma.201003581 (2011).

10  Matthews, J. & Blakeslee, A. Defects in epitaxial multilayers: I. Misfit dislocations. *Journal of Crystal Growth* **27**, 118-125, doi:10.1016/S0022-0248(74)80055-2 (1974).

11  Warren, W. L., Vanheusden, K., Dimos, D., Pike, G. E. & Tuttle, B. A. Oxygen Vacancy Motion in Perovskite Oxides. *Journal of the American Ceramic Society* **79**, 536-538, doi:10.1111/j.1151-2916.1996.tb08162.x (1996).

12  Jiang, W., Noman, M., Lu, Y. M., Bain, J. A., Salvador, P. A. & Skowronski, M. Mobility of oxygen vacancy in $SrTiO_3$ and its implications for oxygen-migration-based resistance switching. *Journal of Applied Physics* **110**, 034509, doi:10.1063/1.3622623 (2011).

13  Muller, D. A., Nakagawa, N., Ohtomo, A., Grazul, J. L. & Hwang, H. Y. Atomic-scale imaging of nanoengineered oxygen vacancy profiles in $SrTiO_3$. *Nature* **430**, 657-661, doi:10.1038/Nature02756 (2004).

14  Nakahara, S. Microporosity Induced by Nucleation and Growth Processes in Crystalline and Non-Crystalline Films. *Thin Solid Films* **45**, 421-432, doi:10.1016/0040-6090(77)90229-2 (1977).

15  Als-Nielsen, J. & McMorrow, D. *Elements of modern X-ray physics.*  (Wiley, 2001).

16  Robinson, I. K. & Tweet, D. J. Surface X-Ray-Diffraction. *Reports on Progress in Physics* **55**, 599-651, doi:10.1088/0034-4885/55/5/002 (1992).

17  Sinha, S. K., Sirota, E. B., Garoff, S. & Stanley, H. B. X-Ray and Neutron-Scattering from Rough Surfaces. *Physical Review B* **38**, 2297-2311, doi:10.1103/Physrevb.38.2297 (1988).

18  Brauer, S., Stephenson, G. B., Sutton, M., Bruning, R., Dufresne, E., Mochrie, S. G. J., Grubel, G., Alsnielsen, J. & Abernathy, D. L. X-Ray-Intensity Fluctuation Spectroscopy Observations of Critical-Dynamics in $Fe_3Al$. *Physical Review Letters* **74**, 2010-2013, doi:10.1103/Physrevlett.74.2010 (1995).

19  Chu, B., Ying, Q. C., Yeh, F. J., Patkowski, A., Steffen, W. & Fischer, E. W. An X-Ray Photon-Correlation Experiment. *Langmuir* **11**, 1419-1421, doi:10.1021/La00005a001 (1995).

20  Dierker, S. B., Pindak, R., Fleming, R. M., Robinson, I. K. & Berman, L. X-Ray Photon-Correlation Spectroscopy Study of Brownian-Motion of Gold Colloids in Glycerol. *Physical Review Letters* **75**, 449-452, doi:10.1103/Physrevlett.75.449 (1995).

21  Mochrie, S. G. J., Mayes, A. M., Sandy, A. R., Sutton, M., Brauer, S., Stephenson, G. B., Abernathy, D. L. & Grubel, G. Dynamics of block copolymer micelles revealed by x-ray intensity fluctuation spectroscopy. *Physical Review Letters* **78**, 1275-1278, doi:10.1103/Physrevlett.78.1275 (1997).

22  Sutton, M., Mochrie, S. G. J., Greytak, T., Nagler, S. E., Berman, L. E., Held, G. A. & Stephenson, G. B. Observation of Speckle by Diffraction with Coherent X-Rays. *Nature* **352**, 608-610, doi:10.1038/352608a0 (1991).

23  Thurn-Albrecht, T., Steffen, W., Patkowski, A., Meier, G., Fischer, E. W., Grubel, G. & Abernathy, D. L. Photon correlation spectroscopy of colloidal palladium using a coherent x-ray beam. *Physical Review Letters* **77**, 5437-5440, doi:10.1103/Physrevlett.77.5437 (1996).

24  Tsui, O. K. C. & Mochrie, S. G. J. Dynamics of concentrated colloidal suspensions probed by x-ray correlation spectroscopy. *Physical Review E* **57**, 2030-2034, doi:10.1103/Physreve.57.2030 (1998).

25  Lumma, D., Lurio, L. B., Mochrie, S. G. J. & Sutton, M. Area detector based photon correlation in the regime of short data batches: Data reduction for dynamic x-ray scattering. *Review of Scientific Instruments* **71**, 3274-3289, doi:10.1063/1.1287637 (2000).

26  Willmott, P. R., Pauli, S. A., Herger, R., Schleputz, C. M., Martoccia, D., Patterson, B. D., Delley, B., Clarke, R., Kumah, D., Cionca, C. & Yacoby, Y. Structural basis for the conducting interface between $LaAlO_3$ and $SrTiO_3$. *Physical Review Letters* **99**, 155502, doi:10.1103/Physrevlett.99.155502 (2007).

27  Zubko, P., Jecklin, N., Torres-Pardo, A., Aguado-Puente, P., Gloter, A., Lichtensteiger, C., Junquera, J., Stephan, O. & Triscone, J. M. Electrostatic Coupling and Local Structural Distortions at Interfaces in Ferroelectric/Paraelectric Superlattices. *Nano Letters* **12**, 2846-2851, doi:10.1021/Nl3003717 (2012).

28   Joyce, B. A. & Joyce, T. B. Basic studies of molecular beam epitaxy - past, present and some future directions. *Journal of Crystal Growth* **264**, 605-619, doi:10.1016/J.Jcrysgro.2003.12.045 (2004).

29   Krug, K., Stettner, J. & Magnussen, O. M. In situ surface X-ray diffraction studies of homoepitaxial electrochemical growth on Au(100). *Physical Review Letters* **96**, 246101, doi:10.1103/Physrevlett.96.246101 (2006).

30   Gutt, C., Ghaderi, T., Chamard, V., Madsen, A., Seydel, T., Tolan, M., Sprung, M., Grubel, G. & Sinha, S. K. Observation of heterodyne mixing in surface x-ray photon correlation spectroscopy experiments (vol 91, art no 076104, 2003). *Physical Review Letters* **91**, 179902, doi:10.1103/Physrevlett.91.179902 (2003).

31   Orsi, D., Cristofolini, L., Baldi, G. & Madsen, A. Heterogeneous and Anisotropic Dynamics of a 2D Gel. *Physical Review Letters* **108**, 105701, doi:10.1103/Physrevlett.108.105701 (2012).

32   Livet, F., Bley, F., Ehrburger-Dolle, F., Morfin, I., Geissler, E. & Sutton, M. X-ray intensity fluctuation spectroscopy by heterodyne detection. *Journal of Synchrotron Radiation* **13**, 453-458, doi:10.1107/S0909049506030044 (2006).

33   Family, F. & Vicsek, T. Scaling of the Active Zone in the Eden Process on Percolation Networks and the Ballistic Deposition Model. *Journal of Physics A - Mathematical and General* **18**, L75-L81, doi:10.1088/0305-4470/18/2/005 (1985).

34  Vicsek, T. & Family, F. Dynamic Scaling for Aggregation of Clusters. *Physical Review Letters* **52**, 1669-1672, doi:10.1103/Physrevlett.52.1669 (1984).

35  Sneppen, K., Krug, J., Jensen, M. H., Jayaprakash, C. & Bohr, T. Dynamic Scaling and Crossover Analysis for the Kuramoto-Sivashinsky Equation. *Physical Review A* **46**, R7351-R7354, doi:10.1103/PhysRevA.46.R7351 (1992).

36  Sutton, M. A review of X-ray intensity fluctuation spectroscopy. *Comptes Rendus Physique* **9**, 657-667, doi:10.1016/J.Crhy.2007.04.008 (2008).

37  Sikharulidze, I., Dolbnya, I. P., Fera, A., Madsen, A., Ostrovskii, B. I. & de Jeu, W. H. Smectic membranes in motion: Approaching the fast limits of X-ray photon correlation spectroscopy. *Physical Review Letters* **88**, 115503, doi:10.1103/Physrevlett.88.115503 (2002).

38  Fahnline, D., Yang, B., Vedam, K., Messier, R. & Pilione, L. Intrinsic Stress in a-Germanium Films Deposited by RF-Magnetron Sputtering. *MRS Online Proceedings Library* **130**, doi:10.1557/PROC-130-355 (1988).

39  Moss, S. C. & Graczyk, J. F. Evidence of Voids Within the As-Deposited Structure of Glassy Silicon. *Physical Review Letters* **23**, 1167-1171, doi:10.1103/PhysRevLett.23.1167 (1969).

40  Thornton, J. A. Influence of apparatus geometry and deposition conditions on the structure and topography of thick sputtered coatings. *Journal of Vacuum Science & Technology* **11**, 666-670, doi:10.1116/1.1312732 (1974).

41  Marsen, B. & Sattler, K. Fullerene-structured nanowires of silicon. *Physical Review B* **60**, 11593-11600, doi:10.1103/Physrevb.60.11593 (1999).

42  Messier, R., Giri, A. P. & Roy, R. A. Revised Structure Zone Model for Thin-Film Physical Structure. *Journal of Vacuum Science & Technology A - Vacuum Surfaces and Films* **2**, 500-503, doi:10.1116/1.572604 (1984).

43  Dirks, A. G. & Leamy, H. J. Columnar Microstructure in Vapor-Deposited Thin-Films. *Thin Solid Films* **47**, 219-233, doi:10.1016/0040-6090(77)90037-2 (1977).

44  Henderson, D., Brodsky, M. H. & Chaudhari, P. Simulation of structural anisotropy and void formation in amorphous thin films. *Applied Physics Letters* **25**, 641-643, doi:10.1063/1.1655341 (1974).

45  Thornton, J. A. The Microstructure of Sputter-Deposited Coatings. *Journal of Vacuum Science & Technology a-Vacuum Surfaces and Films* **4**, 3059-3065, doi:10.1116/1.573628 (1986).

46  Cargill, G. S. Anisotropic Microstructure in Evaporated Amorphous Germanium Films. *Physical Review Letters* **28**, 1372-1375, doi:10.1103/PhysRevLett.28.1372 (1972).

47  Zhou, L., Wang, Y. P., Zhou, H., Li, M. H., Headrick, R. L., MacArthur, K., Shi, B., Conley, R. & Macrander, A. T. Pressure-dependent transition from atoms to nanoparticles in magnetron sputtering: Effect on $WSi_2$ film roughness and stress. *Physical Review B* **82**, 075408, doi:10.1103/Physrevb.82.075408 (2010).